\documentclass[10pt,letterpaper]{article}
\usepackage{opex3}
\usepackage[margin=0.5in,footnotesize,labelsep=space]{caption}
\usepackage[listofformat=parens,subrefformat=subparens]{subfig}
\usepackage{amsmath}
\usepackage{graphics}
\usepackage{epstopdf}
\usepackage{cite}
\usepackage{bm}
\usepackage[section]{placeins}
\usepackage{amssymb}
\usepackage[utf8]{inputenc}
\usepackage[T1]{fontenc}
\usepackage{mathptmx}
\usepackage{microtype}
\usepackage[squaren]{SIunits}
\let\originalleft\left
\let\originalright\right
\renewcommand{\left}{\mathopen{}\mathclose\bgroup\originalleft}
\renewcommand{\right}{\aftergroup\egroup\originalright}
\newcommand{\ket}[1]{\left|#1\right\rangle} 
\newcommand{\exval}[3]{\left\langle#1\middle|#2\middle|#3\right\rangle}

\newcommand{\exprod}[2]{\left|#1\middle\rangle\middle\langle#2\right|}
\newcommand{\sinc}{\operatorname{sinc}} 
\newcommand{\enquote}[1]{``#1''}

\begin{document}
\title{Control of quantum transverse correlations on a four-photon system}
\author{P.-L. de Assis$^{1,*}$, M. A. D. Carvalho$^1$, L. P. Berruezo$^1$, J. Ferraz$^1$, I. F. Santos$^2$, F. Sciarrino$^{3,4}$, S. Pádua$^1$}
\address{$^1$Departamento de Física, Universidade Federal de Minas Gerais,\\ Caixa Postal 702, 30123-980, Belo Horizonte, Brazil\\$^2$Departamento de Física, ICE, Universidade Federal de Juiz de Fora, \\Juiz de Fora, CEP 36036-330, Brazil\\$^3$Dipartimento di Fisica, Sapienza Universitá di Roma, Roma 00185, Italy \\ $^4$Istituto Nazionale di Ottica, Largo Fermi 6, I-50125 Firenze, Italy}
\email{plouis@fisica.ufmg.br}
\begin{abstract}
Control of spatial quantum correlations in bi-photons is one of the fundamental principles of Quantum Imaging. Up to now, experiments have been restricted to controlling the state of a single bi-photon, by using linear optical elements. In this work we demonstrate experimental control of quantum correlations in a four-photon state comprised of two pairs of photons. Our scheme is based on a high-efficiency parametric down-conversion source coupled to a double slit by a variable linear optical setup, in order to obtain spatially encoded qubits. Both entangled and separable pairs have been obtained, by altering experimental parameters. We show how the correlations influence both the interference and diffraction on the double slit.
\end{abstract}

\ocis{(270.0270) Quantum Optics; (270.4180)   Multiphoton processes; (270.5585) Quantum information and processing.}

\section{Introduction}\label{introduction}

Works exploring the quantum effects on multi-photon spatial interference and images have traditionally focused on experiments using a pair of photons \cite{Jaeger:1993p2577,Horne:1989p2576,Ribeiro:1994p4176,Pittman:1996p2565,Abouraddy:2001p2562}. Other important results obtained in the past decades, involving the spatial degree of freedom of photon pairs, were the complementarity between entanglement and single-photon coherence \cite{Saleh:2000p2568}, increased resolution on imaging when using entangled photons \cite{Santos:2005,Brida:2009,Brida:2010}, the spatial quantum eraser \cite{Walborn:2002}, measurement of the two-photon de Broglie wavelength \cite{Fonseca:1999,Fonseca:2001}, and the experiments encoding logical qubits and qudits onto the path state of photons going through slits \cite{Lima:2006p2560,Neves:2004p2556,Neves:2005p2557,Neves:2007p2551,Tagushi:2008,Peeters:2009p219}.

In this paper we demonstrate control over the quantum correlations of two biphotons that are encoded in a four qubit space by using a double slit. Our setup is able to generate two pairs of highly entangled photons in one configuration and four nearly-separable photons in another. We further present a way of characterizing the state generated, without having to resort to fourfold coincidence detection, by exploiting the fact that each pair contains one photon of horizontal polarization and one of vertical polarization. This allows us to observe both the contribution that is common to single and double pair spontaneous parametric down-conversion (SPDC) events and the one that is exclusive to double pair events, i.e. coincidences between photons with the same polarization. Finally, we also observed that the diffraction behavior depended on the quantum correlations of the photons, producing an asymmetrical pattern in the case of entangled photons and the expected diffraction pattern for the case of photons in a nearly-separable spatial state.

By encoding the transverse profile of photons on a discrete space with the use of slits, it becomes simple to scale the dimensionality, from qubits to qudits, by using more slits. This experiment, then, opens a path for future works using multi-pair qudit states. As it is customary, double-slits and multi-slits are characterized by their width $2a$ and by the center-to-center distance $2d$.

\section{The four-photon state}\label{4photon}
In order to exploit the possibility of scaling experiments to include quantum systems of dimensions greater than 2, it has become usual to encode logical qubits and qudits on the transverse momentum and/or position state of photons. Our experiment encodes the logical qubits $\ket{0}$ and $\ket{1}$ on the condition that the photon passes through the left or right slit, respectively.

As presented for the two-photon case in \cite{Neves:2007p2551,Peeters:2009p219}, the most general state of a pair of qubits generated using a double slit and assuming a symmetrical illumination can be written as:
\begin{equation}\label{general}
\ket{\Psi_{H,V}}=\cos(\alpha/2)\ket{\psi^+}+e^{i\varphi}\sin(\alpha/2)\ket{\phi^+}
\end{equation}
where $\ket{\psi^+}$ and $\ket{\phi^+}$ are maximally entangled states of two photons A and B known as Bell states:
\begin{eqnarray}
\ket{\psi^+}=\frac{1}{\sqrt{2}}\left(\ket{01}_{A,B}+\ket{10}_{A,B}\right)\\
\ket{\phi^+}=\frac{1}{\sqrt{2}}\left(\ket{00}_{A,B}+\ket{11}_{A,B}\right)
\end{eqnarray}

Up to now, experiments have been restricted to controlling the spatial state of a single bi-photon, by using linear optical elements. In this work we demonstrate experimental control of quantum correlations in a four-photon state comprised of two pairs of photons. By using a pulsed pump and a high-efficiency periodically-poled KTiOP$\mathrm{O}_4$ crystal \cite{Armstrong:1962p1462, Wong:1p2578, Kuklewicz:2004p446}, the second order term in $\chi^{(2)}$ on the expansion of the time-evolution operator associated to the SPDC becomes relevant. In our experimental conditions, this term results in the generation of two distinguishable pairs of photons.

The pump is a \unit{413}{nm} pulsed beam obtained by second harmonic generation on a BBO crystal pumped by an \unit{826}{nm} femtosecond pulsed laser that has pulses of 200 fs and a repetition rate of \unit{76}{MHz}. The mean power of the pump beam when falling on the crystal is of the order of \unit{80}{mW}. It is loosely focused by a \unit{30}{cm} spherical lens onto a \unit{10}{mm} long, type-II PPKTP grown for collinear and degenerate SPDC centered on $\lambda=\unit{826}{nm}$. The created pair will consist of one horizontally (H) and one vertically (V) polarized photon. Polarization was used as an auxiliary mode, in order to separate the photons of each pair into two distinct physical subsystems A and B, much in the same way non-collinear experiments that encode the qubits using polarization use the down-conversion linear momentum modes. 

The general state, considering one or two pairs cross the double-slit is, up to appropriate normalization \cite{Ou:book},
\begin{eqnarray}
\ket{\Psi}=M\left(\ket{vac}+\eta\ket{\Psi_{H,V}}+\frac{\eta^2}{2}\ket{\Psi_{H,V}}_I\otimes\ket{\Psi_{H,V}}_{II}\right)
\end{eqnarray}
where $\eta$ is the generation efficiency, depending on both the non-linear susceptibility and the pump intensity, and $M$ is a normalization constant.

The four-photon component can then be written as
\begin{eqnarray}
\ket{\Psi_{H,V}}_1\otimes\ket{\Psi_{H,V}}_2=\left(\cos(\alpha/2)\ket{\psi^+}+e^{i\varphi}\sin(\alpha/2)\ket{\phi^+}\right)^{\otimes2}
\end{eqnarray}

Since there is no genuine multipartite entanglement, it is sufficient to characterize the entanglement by the concurrence \cite{Wootters:1998p445} of the single pair, which is given by \cite{Neves:2007p2551,Peeters:2009p219}
\begin{equation}\label{concurrence}
\mathcal{C}=\sqrt{1-(\cos(\varphi)\sin(\alpha))^2}
\end{equation}

States with $\varphi=\pi/2$ or $\alpha=0,\pi$ are maximally entangled. Conversely, states with $\varphi=0$ and $\alpha=\pi/2$ are separable.

The presence of two photons of horizontal and vertical polarization also gives us the opportunity to consider the partial state of same-polarization photons. In practical terms, this is obtained by using a polarized beamsplitter to create two detection branches, one for horizontally and one for vertically polarized photons.

Since the two pairs are not entangled, the purity of the partial state, defined as $\mathrm{Tr}(\rho^2_{partial})$ and given below, will be used as a relevant indicator on those same-polarization coincidences, as it is directly related to the degree of entanglement in each pair:
\begin{equation}\label{purity}
\mathcal{P}=\frac{1}{4}\left(1+(\cos(\varphi)\sin(\alpha))^2\right)^2.
\end{equation}
The combination of Eq. (\ref{concurrence}) and Eq. (\ref{purity}) allows us to write
\begin{equation}
\mathcal{P}=\frac{1}{4}(2-\mathcal{C}^2)^2.
\end{equation}

\section{Control and characterization of spatial quantum correlations}\label{control}

In order to control the quantum correlations in the four photons, we will resort to a tunable linear optical scheme. Here we present two extremal conditions, that is, the pairs of photons coming out of the double slit are either highly entangled or separable. The correlations after the double-slit depend on how the bi-photon field generated inside the crystal is projected onto the plane of the apertures. Influence over the state encoded on the slits can be exerted either by controlling the pump field, the propagation of the bi-photons from the crystal to the slits or both. Since the bi-photon field is not affected in one of the transverse dimensions by the double-slit (the larger dimension, i.e. the $y$ direction), we restrict our analysis to the $x$ direction (the smallest slit dimension).

More specifically, the two-photon field in the crystal can be written, in the transverse momentum variables, as \cite{Monken:1998p2559,Peeters:2009p219}
\begin{equation}
\tilde{\Phi}(q_1,q_2)=\tilde{E}_p(q_1+q_2)\tilde{\xi}(q_1-q_2),
\end{equation}where $\{q_1,q_2\}$ are the transverse components of the momentum of photon 1 and 2 in the bi-photon. The tilde indicates that the functions refer to the transverse momentum space. Therefore $\tilde{E}_p$ is the angular spectrum of the pump beam and $\tilde{\xi}$ is the phase-matching condition for the PPKTP.

In most cases, including our experiment, $\tilde{E}_p$ is the angular spectrum of a gaussian beam. Hence, $\tilde{E}_p(q_1+q_2)\propto\mathcal{F}(\exp(-(\frac{x_1+x_2}{2w_p})^ 2))$ when the beam waist, $w_p$, is in the center of the crystal, with $\mathcal{F}$ corresponding to the Fourier transform and $x_1$ and $x_2$ to the transverse position of photons 1 and 2. If the last condition is not satisfied, phase components due to the curvature of the wavefront must be considered as well.

For our experimental conditions, the phase matching function is 
\begin{equation}
\tilde{\xi}\propto\sinc\left(\phi_0+\frac{L(q_1-q_2)^2}{(8n_{eff}\omega_{SPDC}/c)}\right)\end{equation} where $\phi_0$ is a phase-mismatch term that we consider to be negligible, $\omega_{SPDC}$ is the frequency of the degenerate down-converted photons and the effective refractive index $n_{eff}=2(n_1n_2)/(n_1+n_2)$ takes into account the fact that the two photons in type-II SPDC propagate differently inside a crystal of length L. A more thorough description of the phase-matching function can be found in \cite{Cosme:arxiv,Peeters:2009p219}.

\begin{figure}[htbp]
\centering
\subfloat[Crystal image configuration]{\label{position}\includegraphics[width=0.75\textwidth]{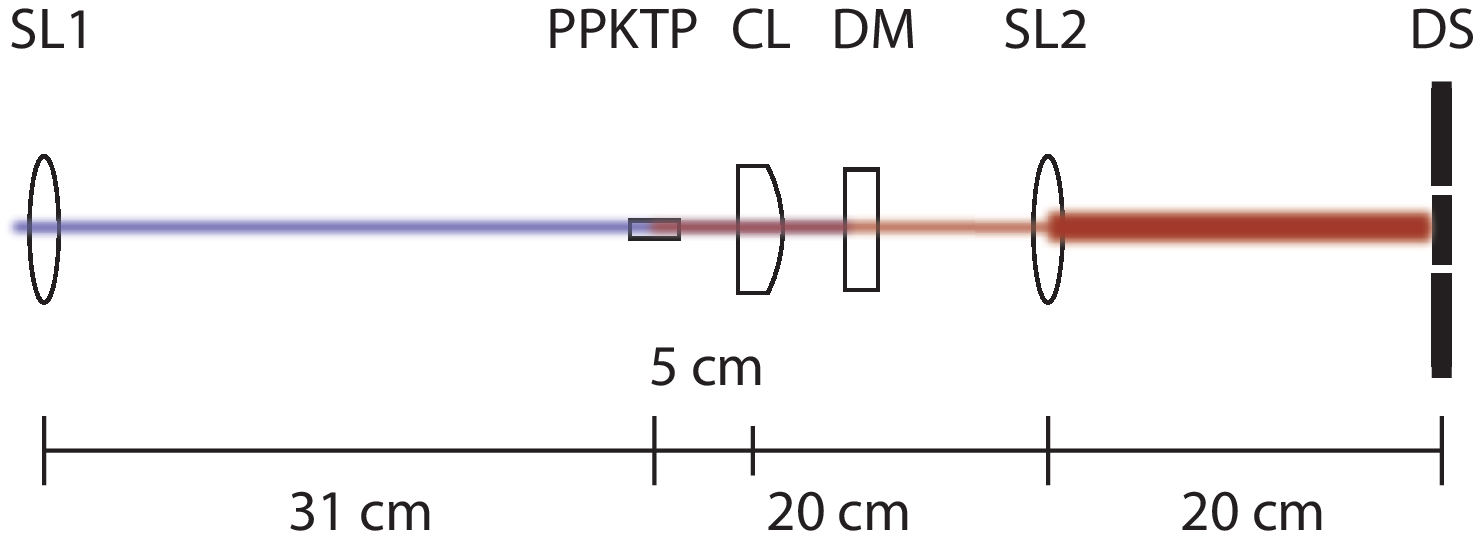}}\\                
\subfloat[Crystal far-field configuration]{\label{momentum}\includegraphics[width=0.75\textwidth]{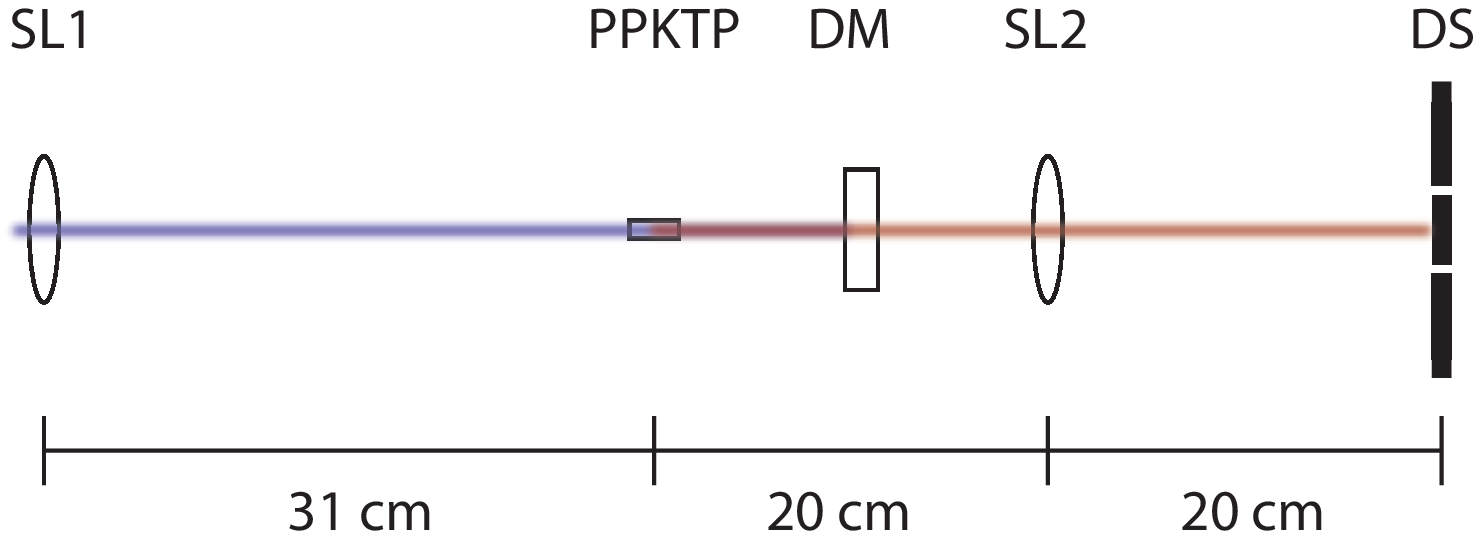}}
\caption{Experimental setup for control of transverse correlations over the double slit (DS). Coupling is done so that either the position \textbf{(a)} or momentum \textbf{(b)} degree of freedom of the bi-photons is mapped onto the double slit plane \cite{Goodman:book}. The first uses a cylindrical lens (CL) of $f_1=\unit{5}{cm}$ and a spherical lens (SL2) of $f_2=\unit{20}{cm}$ to project a magnified image of the crystal center, while the latter uses only the $f_2=\unit{20}{cm}$ lens to project the Fourier transform of the same plane onto the plane of the double slit. The slits are \unit{80}{\micro\metre} wide and have a separation $2d=\unit{240}{\micro\metre}$. SL1 is the $f=\unit{31}{cm}$ spherical lens that focuses the pump beam on the PPKTP crystal and DM is a dichroic mirror that reflects the \unit{413}{nm} beam while allowing the \unit{826}{nm} down-converted photons to pass.}
\label{setup}
\end{figure}

Figure \ref{setup} shows the experimental setups used in this work to control the quantum correlations of the bi-photons at the double-slit plane. We adopted two configurations: \subref{position} to generate highly entangled states and \subref{momentum} to prepare separable states. 

Scheme \subref{position} projects a magnified image of the crystal plane onto the double-slit. This geometry is achieved by adopting two lenses, a cylinder lens with focal lens $f_1=\unit{5}{cm}$ and a spherical one with $f_2=\unit{20}{cm}$, so the magnification factor of the image reads $M=f_2/f_1=4$. We adopt a cylinder lens in order to create a crystal image in the direction perpendicular to the slits, the x-direction, while keeping the far-field image in the y orthogonal direction. By such approach the dominant contribution to the bi-photon amplitude transmitted by the double-slit is the one associated to the phase-matching function. That is so because, in the magnified image of the crystal center projected on the slits, the gaussian profile of the pump component is much wider than the Fourier transform of the $\sinc$ function of the phase-matching condition and can be considered constant.

Configuration \subref{momentum} relies on the far-field imaging onto the double slit. This scheme is obtained by adopting a single spherical lens with focal length $f=\unit{20}{cm}$ placed at a distance $f$ from both the center of the crystal and the plane of the slits. With this geometry the bi-photon amplitude is almost constant along the double-slit apertures leading to the generation of separable states.

A detailed discussion of the relation between the parameters $\alpha$ and $\varphi$ of Eq. (\ref{general}) and the adopted geometries can be found in \cite{Peeters:2009p219}, in the form of what was called an engineering parameter $p$:\begin{align}p=\exp(i\varphi)\tan(\alpha/2)=\frac{A(d,d)}{A(d,-d)},\end{align} that gives a clear method for estimating the values of $\alpha$ and $\varphi$ by knowing the form of the two-photon amplitude function $A(x_1,x_2)$ specifically when $x_1=x_2=d$ and $x_1=-x_2=d$, where $x_1$ and $x_2$ are coordinates on the plane of the double-slit and $d$, as previously mentioned, is the distance from the center of one of the slits to the center of the slit pattern. In this case, the distance between the center of the slits is $2d$. We refer to the values calculated through that method when giving expected values for $\alpha$ and $\varphi$. For our experimental setups, the expected values are $\alpha_a=\unit{173}{\degree}$ and $\varphi_a=\unit{180}{\degree}$ in the configuration of Fig. \subref{position}. By changing to the configuration of Fig. \subref{momentum} we expect $\alpha_b=\unit{85}{\degree}$ and $\varphi_b=\unit{0}{\degree}$.

Let us now discuss the measurements necessary to characterize the states generated by the setups presented in Fig. \ref{setup}.

The parameter $\alpha$ can be inferred by measurements on the image of the slits, which corresponds to the $\left\{\ket{0},\ket{1}\right\}$ basis. Measurements for determining $\varphi$, however, require us to go to the diagonal basis. For qubits encoded on transverse spatial variables, this can be achieved by looking at coincidence measurements on the pattern generated in the far field of the slits, that gives information on both $\alpha$ and $\varphi$. One way to access the far field is to use a lens with focal length $f_{FL}$ in the $f$-$f$ configuration. This maps the Fourier transform of the slits onto positions on the detection plane, as shown in Eq. 5-19 of \cite{Goodman:book}.

The probability of coincident detection of two photons from the same pair, one of horizontal polarization and the other of vertical polarization, when using the far-field strategy, can be derived as a function of the position of both detectors involved, $P(x_1,x_2)$. To calculate it, we first assume the contribution of only a single pair of photons and then follow Mandel and Wolf \cite{Mandel:book} to obtain
\begin{equation}
P(x_1,x_2)_{(H,V)}\propto\left|E^+_H(x_1)E^+_V(x_2)\ket{\Psi_{H,V}}\right|^2,
\end{equation}
where $E^+_j(x_i)$, for $j=(H,V)$ and $i=(1,2)$ is calculated by propagating the electrical field operators from the plane of the slits to the detection plane, taking into account the lens placed at a distance $f_{FL}$ from both planes. This, as mentioned before, means taking the optical Fourier transform of the product of the aperture function and the amplitude of the bi-photon on the plane of the slits \cite{Neves:2005p2557,Peeters:2009p219}.

After performing the calculations, we obtain the explicit form of  $P(x_1,x_2)_{(H,V)}$:
\begin{align}
P(x_1,x_2)_{(H,V)}  \propto \; &1+\cos ^2(\alpha/2)\cos(\beta(x_1-x_2)) + \nonumber \\
&|\sin(\alpha)|\cos(\varphi)\left[\cos(\beta(x_1))+\cos(\beta(x_2))\right] + \nonumber \\
&\sin ^2(\alpha/2)\cos(\beta(x_1+x_2)).\label{pDBC}
\end{align} 

This function can be represented on a plane, which we will call the coincidence map. The same approach can be adopted to represent the coincidence counts detected in the experiment, allowing us to compare experimental and theoretical maps. We see that it depends of both $\alpha$ and $\varphi$, meaning the coincidence maps can be used to characterize both parameters of the state.

The following step is to characterize the measurements sensitive exclusively to the two-pair contribution $\ket{\Psi_{H,V}}_I\otimes\ket{\Psi_{H,V}}_{II}$.

\begin{figure}[htbp]
\centering
\includegraphics[width=0.60\textwidth]{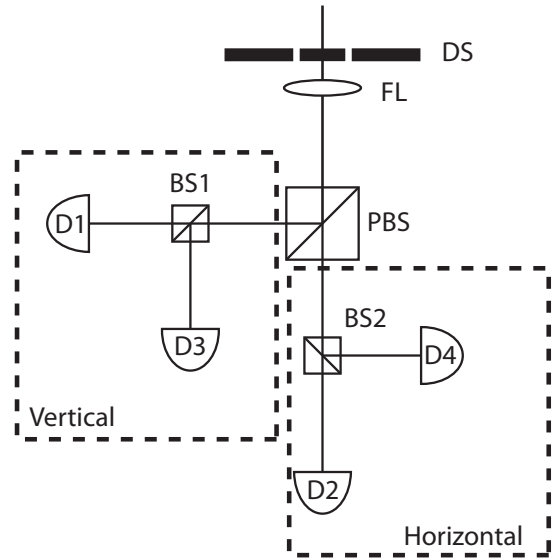}
\caption{A schematic view of the detection apparatus, showing how the incoming photons are divided into two branches by a PBS. On each branch the photons are further divided by a BS. This allows us to make coincidence detection between two detectors on the same branch or between detectors on different branches. Detections on the same branch originate only from double-pair events, while the raw different branch detections are a sum of contributions from single-pair and double-pair events. In this figure, the focal plane of FL corresponds to the detection plane, upon which all four detectors are free to move parallel to the plane of the table and perpendicular to the direction of beam propagation after passing through the beamsplitters. DS is the double-slit.}
\label{detection}
\end{figure}

Since we work with two pairs of photons, the most natural type of measurement would be fourfold coincidence detection. This, however, is impractical due to the fact that only a small fraction of the total signal is collected by the detectors when in the far-field regime. As fourfold coincidences depend on the overall efficiency of four detectors, it can require extremely long detection times. To circumvent this predicament, we use the fact that our source of photon pairs is a type-II crystal and, thus, coincidences on the same polarization branch can only come from down-conversion events in which two pairs were created. Another factor that favors same-branch coincidences as a method for investigating the two-pair signature is that only one beamsplitter (BS) is involved, whereas two are necessary on fourfold detection events, as can be seen in Fig. \ref{detection}. 

These coincidences can be mapped in much the same way as those from two photons with orthogonal polarization. Even though double-pair events are not the major contribution, the probability of detecting H-H or V-V coincidences is still much larger than of detecting fourfold coincidences, since only two detectors are involved in the process and, therefore, only two detection efficiencies are taken into account.

The expression for coincidence detection of photons on the same polarization branch is:
\begin{eqnarray}
P(x_1,x_2)_{\{(H,H);(V,V)\}}\propto 1+(\sin(\alpha)\cos(\varphi))^2\cos(x_1)\cos(x_2)+\nonumber\\
\sin(\alpha)\cos(\varphi)(\cos(x_1)+\cos(x_2)).\label{pSBC}
\end{eqnarray}

To calculate it, we use
\begin{eqnarray}\label{partialprob}
P(x_1,x_2)_{\{(H,H);(V,V)\}}\propto \exval{vac}{E^+(x_1)\rho^\prime_{(I,j)}E^-(x_1)}{vac}\times \nonumber \\ \exval{vac}{E^+(x_2)\rho^\prime_{(II,j)}E^-(x_2)}{vac},
\end{eqnarray} in which $j$ can be either $H$ or $V$. Indexes $I$ and $II$ correspond to the two pairs of photons. The density matrix $\rho^\prime_j$ is given by $\rho^\prime_j=\mathrm{Tr}_i\left(\exprod{\Psi_{HV}}{\Psi_{HV}}\right)$, with $\mathrm{Tr}_i$ being the partial trace over the photon of polarization $i\neq j$. Note that the same derivation can be used to calculate the probability of detecting coincidences between photons of different polarization and different pairs.

Equations (\ref{pDBC}) and (\ref{pSBC}) have been derived under the assumption that the incident two-photon field can be considered constant over the width of a single slit, which is taken to be very narrow. This allowed us to factor out the diffraction term as an enveloping function that is radially symmetric around the origin in the plane formed by $x_1$ and $x_2$. As we will discuss in Section \ref{diffraction}, this is not true for all situations, but is a valid approximation if one is interested only in studying, for instance, the interference behavior close to the center of the detection plane.

We will now discuss in detail the experimental procedure used to obtain the coincidence maps. The detection setup, shown in Fig. \ref{detection} is composed of a spherical lens FL of focus length $f=\unit{30}{cm}$ and four avalanche photo-diodes (APD) detectors divided into two polarization branches, as indicated above by the dashed bounding boxes. Each branch corresponds to one of the outputs of a polarized beamsplitter (PBS) and contains a balanced beamsplitter (BS1 and BS2). Thus, coincidences can be detected between APDs on the same polarization branch, to which we refer as \textbf{SBC} (\textbf{S}ame \textbf{B}ranch \textbf{C}oincidences), and on different branches, referred to as \textbf{DBC} (\textbf{D}ifferent \textbf{B}ranch \textbf{C}oincidences). Each detector is mounted on a manual translation stage in the x direction, defined as being parallel to the optical table plane and orthogonal to the propagation direction of the beam. Additionally, a \unit{200}{\micro\metre} pinhole and an interference filter ($\Delta\lambda=\unit{40}{nm}$) centered on \unit{826}{nm} are placed in front of the APD. 

Coincidences are detected in a 1 ns window using a custom electronic circuit. Such a time frame is large enough to collect all events originated from a single \unit{200}{fs} pulse but small enough so that there can be no coincidence between photons from two different pulses, allowing us to consider each pulse as an isolated SPDC event that may generate either zero, one or two pairs of photons. The maps are the result of 21 scans (17 in the case of Fig. \ref{entangled_diffraction}) where two detectors, one in each branch remain in a fixed position and the other two are moved in steps of \unit{100}{\micro\metre} (for fine scans over the main interference region) or \unit{500}{\micro\metre} for broad scans, meant for observing the diffraction pattern over a larger area.

In addition to generating the maps, the data collected was also used to evaluate efficiency of our source. To calculate the contribution of single and double pair events, we have estimated the detection efficiency of our apparatus starting from the measured count and coincidence rates of detectors. Considering uncorrelated spatial photon pairs generated with an average pair number per pulse $p$, detected with an overall efficiency $\gamma$, for an experiment with time duration $\tau$ and repetition rate $R$, the average counts of the detector are equal to $N_1=\gamma pR\tau$, while the measured  coincidences are $N_{12}=\gamma^2 pR\tau$. The value of $\gamma$ can then be obtained as $N_{12}/N_{1}$. Spatial correlations between the two photons can modify the previous relation, so the data used corresponds to the maxima of detection on each map.

\section{Experimental results}\label{results}

Following the experimental procedures detailed in the previous section, we obtain coincidence maps on the far field of the double slits, which will serve as fingerprints for identifying the states generated. The parameters obtained from these maps were then compared to those predicted by the theoretical model of our experiment, as presented in section \ref{control}.

Maps on Fig. \ref{correlated}, \ref{uncorrelated} and \ref{diffraction_separable}  were generated from a matrix of 21x21 experimental points. Those on Fig. \ref{entangled_diffraction} originate from a 17x17 grid, although the separation between points are the same as that of Fig. \ref{diffraction_separable}, meaning that a smaller field of view was imaged, from -4 to \unit{4}{mm} instead of -5 to \unit{5}{mm}. Data points have been normalized relative to the maximal values of each map, for ease of comparison to the simulated maps based on the theoretical model from Eq. (\ref{pDBC}) and Eq. (\ref{pSBC}). Additionally, the latter are generated over the same grids of points, corresponding to the same values of $x_1$ and $x_2$. This means that artifacts due to interpolation are not exclusive to the experimental maps and should present themselves in the same manner on the simulations, minimizing discrepancies that may influence the interpretation of the processed data. There are, however, some artifacts present only on experimental maps, due to noise on the data set. On the case of maps on Fig. \ref{correlated} and Fig. \ref{uncorrelated}, the dominant aspect is noise, since the structures are much larger than the spacing on the grid. On Fig. \ref{entangled_diffraction} and Fig. \ref{diffraction_separable}, however, the size of the finer structures is comparable to that of the spacing between points on the grid. This makes the aliasing contribution more significant. However, in those two figures the most interesting structures, those due to diffraction, are larger than the grid spacing.

Equation (\ref{pDBC}) was used to simulate the different branch coincidence map for comparing with the experimental DBC map. Since this equation doesn't take in account coincidences between photons of orthogonal polarization but from different pairs, we have subtracted that contribution from $N_{12}$  by estimating the accidental coincidences as $N_{acc}=N_1*N_2/(R\tau)$. These accidental coincidences, calculated for each measured point, were also subtracted from the maps. 

Using the corrected data, we also characterize our source. Calculating the average pair number $p=N_1/(\gamma R\tau)$ we obtain $p_a=0.49$ and $p_b=0.96$, respectively for the setup of Fig. \subref{position} and \subref{momentum}. We attribute the difference between the two values to the different geometries and to the spatial correlation observed with setup \subref{position}. The effective single pass parametric gain, calculated after the double slit,  is 0.65 for setup \subref{position} and 0.87 for \subref{momentum}. The probabilities $P(n)$ of generating $n$ pairs are, for  \subref{position} and \subref{momentum}, $P(1)_a=0.29$, $P(1)_b=0.37$, $P(2)_a=0.07$, $P(2)_b=0.19$, $P(3)_a=0.011$ and $P(3)_b=0.066$.

We now examine in detail the results of the entangled state obtained by using the setup of Fig. \subref{position} and, after it, the near-separable state generated by the setup indicated in \subref{momentum}.

\subsection{Entangled state}

\begin{figure}[htb]
\centering\includegraphics[width=\textwidth]{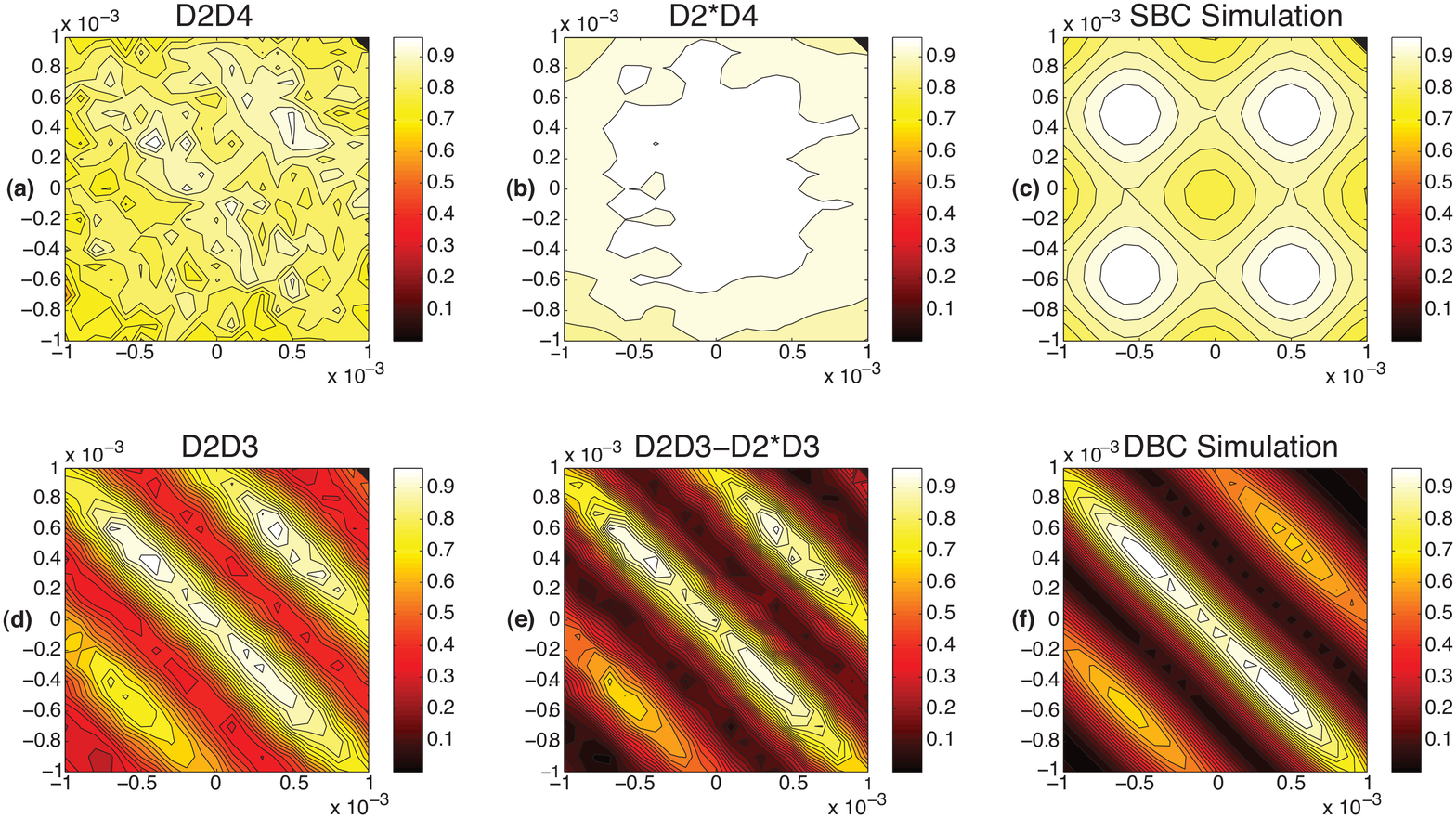}
\caption{Normalized two-photon measured and simulated coincidence maps for a four-photon path state presenting a high degree of spatial correlation in each bi-photon. The upper row, from \textbf{(a)} to \textbf{(c)}, includes the measured coincidences on the H branch (D2D4), the calculated uncorrelated pattern corresponding to the product of the single counts (D2*D4), weighted by the repetition rate of the pump, and the SBC simulation, generated with Eq. (\ref{pSBC}). The lower row, from \textbf{(d)} to \textbf{(f)}, corresponds respectively to the coincidences between the H and V branches (D2D3) and to the same map corrected by subtracting the uncorrelated contribution (D2D3-D2*D3), as well as a simulated DBC map, generated with Eq. (\ref{pDBC}). The simulation parameters used are $\alpha_a^{exp}=\unit{176}{\degree}$ and $\varphi_a^{exp}=\unit{170}{\degree}$, corresponding to $\mathcal{C}\approx0.997\pm0.003$. For a measurement time of \unit{25}{s} per point, the maximum in map \textbf{(a)} corresponds to 549 coincidence counts and in map \textbf{(d)} to 1630 coincidence counts. The measured maps were obtained at the Fourier plane of Fig. \ref{detection}, after the photon pairs cross the double-slit.}
\label{correlated}
\end{figure}

In Fig. \ref{correlated}, the SBC map, originated only by multi-pair events, confirms that the partial state is indeed mixed. The fact that a small degree of structure, still present on the simulated SBC map, is not observed experimentally is attributed to detection fluctuations that mask the low visibility pattern. The interference pattern observed on experimental DBC maps, which reflect mainly the single-pair contribution, is consistent with near maximal entanglement on the qubit picture. Those two results are interpreted as a sign that two distinct pairs of photonic spatial qubits with high purity and path entanglement have been generated.

In Fig. \ref{correlated}(d) and Fig. \ref{correlated}(e) one can notice that the visibility increases substantially after subtracting from the total detections the coincidences between photons of different pairs generated by the same pulse, and transmitted through the double-slit. Our result then demonstrates that highly entangled pairs can be generated even in the regime where many bi-photons are produced by a single pump pulse. We also note that, on each of the three fringes visible in the pattern, the maxima are situated on $x_1=x_2=\pm$\unit{500}{\micro\meter} or \unit{0}{\micro\meter}. This period of \unit{500}{\micro\meter} coincides with the step used on the measurements of Fig. \ref{entangled_diffraction}.

As previously stated, the expected values for the two characteristic parameters for the state were $\alpha_a=\unit{173}{\degree}$ and $\varphi_a=\unit{180}{\degree}$. In order to maximize the agreement between the experimental data and the simulation, those two parameters were changed to $\alpha_a^{exp}=\unit{176}{\degree}$ and $\varphi_a^{exp}=\unit{170}{\degree}$. Such a change for $\varphi$ would be expected if the position of the pump waist and of the focal plane of the cyllindrical lens differed by approximately 200 microns, which is well within the experimental uncertainties of our setup. For the values used in the simulation, the concurrence is $\mathcal{C}\approx0.997\pm0.003$ and the purity for the partial state is $\mathcal{P}\approx0.254\pm0.003$, compared to a minimum of 0.25 for a maximally mixed state.

\subsection{Near-separable state}

\begin{figure}[ht]
\centering\includegraphics[width=\textwidth]{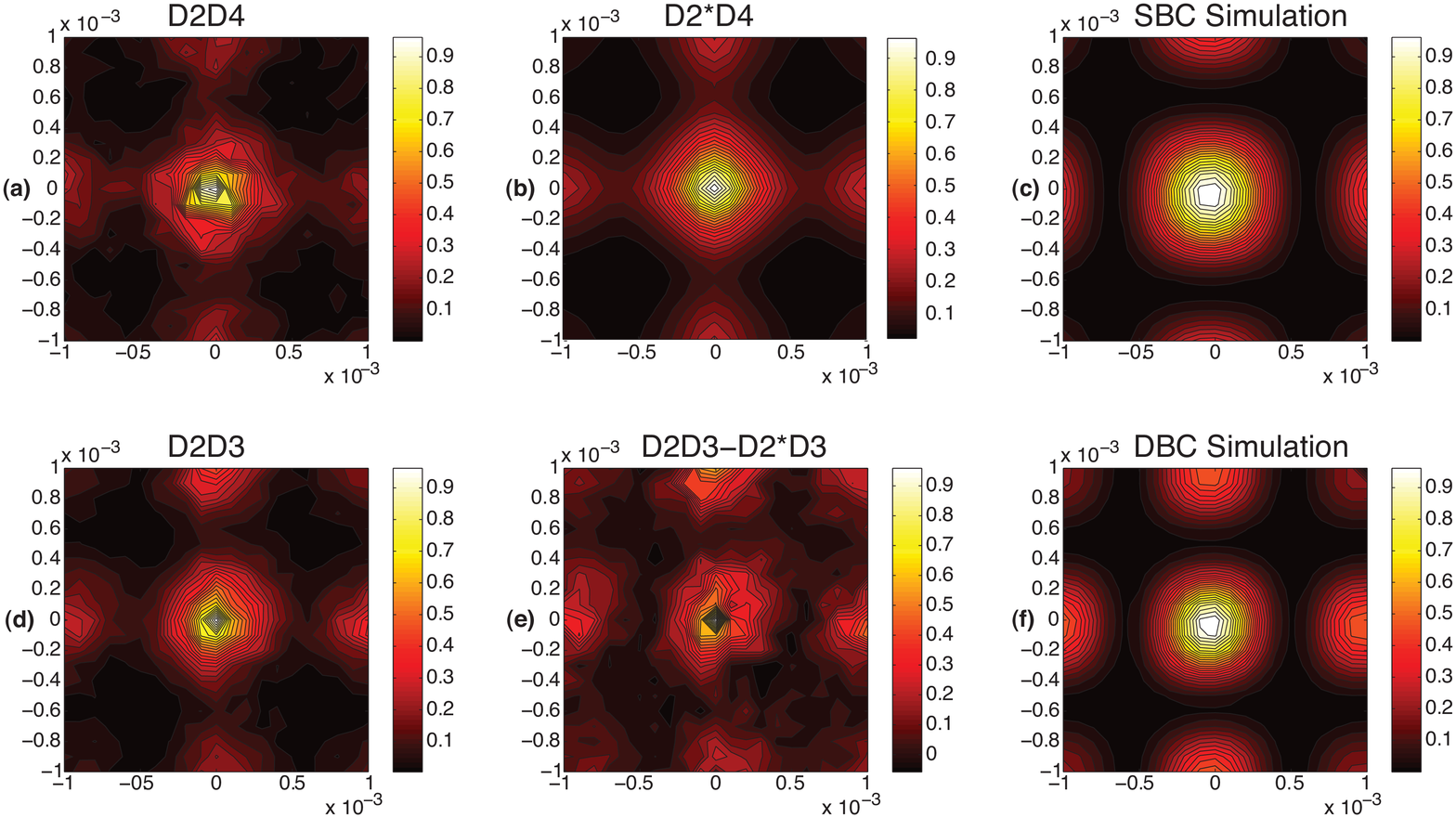}
\caption{Normalized two-photon measured and simulated coincidence maps for a four-photon path state of negligible spatial correlation. The upper row, from \textbf{(a)} to \textbf{(c)}, includes the measured coincidences on the H branch (D2D4), the calculated uncorrelated pattern corresponding to the product of the single counts (D2*D4), weighted by the repetition rate of the pump, and the SBC simulation, generated with Eq. (\ref{pSBC}). The lower row, from \textbf{(d)} to \textbf{(f)}, corresponds respectively to the coincidences between the H and V branches (D2D3) and to the same map corrected by subtracting the uncorrelated contribution (D2D3-D2*D3), as well as a simulated DBC map, generated with Eq. (\ref{pDBC}). The simulation parameters used are $\alpha_b^{exp}=\unit{86}{\degree}$ and $\varphi_b^{exp}=\unit{5}{\degree}$, corresponding to $\mathcal{C}\approx0.11\pm0.03$. For a measurement time of \unit{25}{s} per point, the maximum in map \textbf{(a)} corresponds to 122 counts and in map \textbf{(d)} to 247 counts. The measured maps were obtained at the Fourier plane of Fig. \ref{detection}, after the photon pairs cross the double-slit.}
\label{uncorrelated}
\end{figure}

Let us now analyze the case of the two-photon coincidence maps generated by the setup of Fig. \subref{momentum}, for which the theoretical prediction is a four-photon state of negligible entanglement in the path variables, after the double-slit. In this case, the single- and double-pair contributions can be considered as a tensor product of two and four distinguishable single-photon pure path states, respectively. This is evidenced by the fact that the experimental maps for SBC and DBC are similar. For a state of two pure and separable pairs, the partial state detected will also be pure. Conversely, for maximally entangled path states the partial state is maximally mixed. Thus, we see no interference on the map of Fig. \ref{correlated}(a), but the maps of Fig. \ref{uncorrelated}(a) and Fig. \ref{uncorrelated}(d) are equal. This is in direct relation to what was shown in \cite{Saleh:2000p2568,Abouraddy:2001p1710}.

For this state, the expected values of the two characteristic parameters were $\alpha_b=\unit{85}{\degree}$ and $\varphi_b=\unit{0}{\degree}$. As done during our analysis of the entangled case, those two parameters were changed to $\alpha_b^{exp}=\unit{86}{\degree}$ and $\varphi_b^{exp}=\unit{5}{\degree}$ in the simulated maps. Such a change for both $\alpha$ and $\varphi$ is, again, well within the experimental uncertainties of our setup. For the values used in the simulation, the concurrence is $\mathcal{C}\approx0.11\pm0.03$, which results in a high purity for the partial state, $\mathcal{P}\approx0.988\pm0.004$.

One should note that the fluctuations observed in Fig. \ref{uncorrelated} are due to the lower statistics, compared to those of Fig. \ref{correlated}. In the former, Poissonian fluctuations are more significant. Another important point to be observed about the maps in Fig. \ref{uncorrelated} is that the peaks are \unit{1}{mm} away from each other, twice the step used on Fig. \ref{diffraction_separable}.

\section{Diffraction}\label{diffraction}

In order to fully understand our experiment, we now turn our attention briefly to the subject of the diffraction envelope. We present a qualitative discussion of its characteristics for the two extremal cases of transverse spatial quantum correlations that were generated in this work. As in the previous section, we find it appropriate to discuss the case of a highly entangled state and of a near-separable state separately.

\subsection{Entangled state}

\begin{figure}[htbp]
\centering\includegraphics[width=\textwidth]{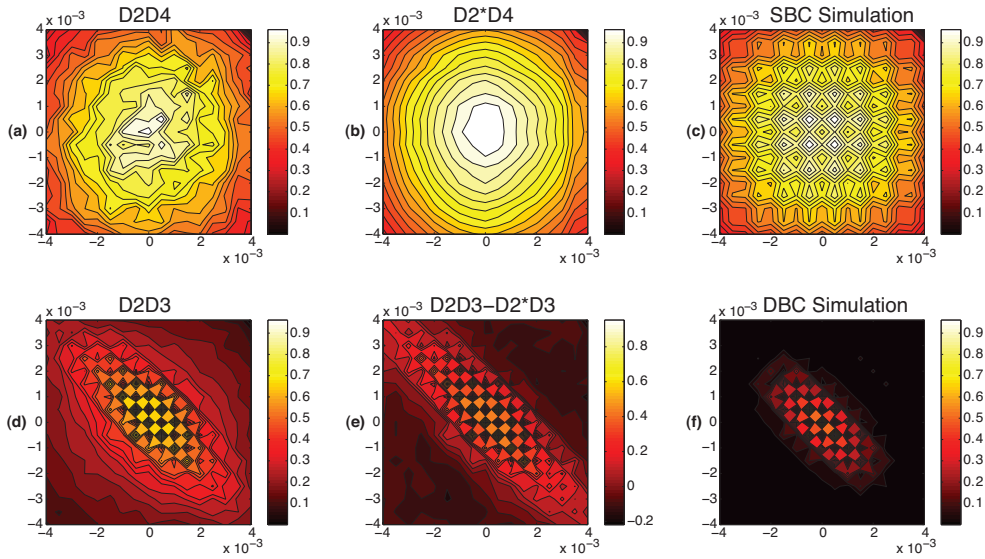}
\caption{Normalized two-photon measured and simulated broad scan coincidence maps for a four-photon spatial state presenting a high degree of spatial correlation in each bi-photon. The upper row, from \textbf{(a)} to \textbf{(c)}, includes the measured coincidences on the H branch (D2D4), the calculated uncorrelated pattern corresponding to the product of the single counts (D2*D4), weighted by the repetition rate of the pump, and the SBC simulation, using Eq. (\ref{pSBC}) and Eq. (\ref{sincpm}). The lower row, from \textbf{(d)} to \textbf{(f)}, corresponds respectively to the coincidences between the H and V branches (D2D3) and to the same map corrected by subtracting the uncorrelated contribution (D2D3-D2*D3), as well as a simulated DBC map, using Eq. (\ref{pDBC}) and Eq. (\ref{sincpm}). The simulation parameters used are, as before, $\alpha_a^{exp}=\unit{176}{\degree}$ and $\varphi_a^{exp}=\unit{170}{\degree}$. For a measurement time of \unit{25}{s} per point, the maximum in map \textbf{(a)} corresponds to 483 counts and in map \textbf{(d)} to 1598 counts.}
\label{entangled_diffraction}
\end{figure}

Figure \ref{entangled_diffraction} shows a larger area for the same experimental conditions as Fig. \ref{correlated}. The asymmetric nature of the diffraction envelope for different branch coincidences is clear, with a larger diffraction on the $(x_1=-x_2)=x_-$ direction and a narrower one for $(x_1=x_2)=x_+$, if compared to the expected homogeneous envelope from previous models concerning diffraction for the same type of experiment \cite{Neves:2007p2551,Peeters:2009p219}. Moreover, the diffraction on the case of same branch coincidences is even larger than predicted by models that consider the amplitude to be constant over a slit.

With the purpose of understanding this, we simulated the diffraction through a single slit for the setup of Fig. \subref{position}, by calculating numerically the optical Fourier transform of the product of the slit transmission function and the amplitude of the bi-photon. This was done on the $x_+$ and $x_-$ directions, which are those that exhibit interesting properties for the different branch coincidences. We show below the optical Fourier transforms used, based once more on Eq. 5-19 of \cite{Goodman:book}:
\begin{equation}
\sinc^2(a_\pm x_\pm)\propto\iint_{x_1=\pm x_2} A(\xi_1,\xi_2)T(\xi_1,\xi_2)\exp\left[-i\frac{k}{f_{FF}}(\xi_1x_1+\xi_2x_2)\right]dx_1dx_2,\label{sincpm}
\end{equation}
where $k$ is the wave number of the down-converted photons, $\xi_1,\xi_2$ are coordinates on the plane of the slits and $x_1, x_2$ are the coordinates on the detection plane. $A(\xi_1,\xi_2)$ is the two-photon amplitude function on the double-slit plane and $T(\xi_1,\xi_2)=T(\xi_1)T(\xi_2)$ is the transmission function for a slit of width $w$, so that $T(\xi_1,\xi_2)=0$ if $|\xi_1|$ or $|\xi_2|>w/2$ and is equal to 1 otherwise. The integrals of Eq. (\ref{sincpm}) were fit by the $\sinc^2$ functions and the parameters $a_+$ and $a_-$ were extracted.

The resulting phenomenological parameters, $a_+$ and $a_-$, were used as input for the simulated maps presented in Fig. \ref{entangled_diffraction}(c) and Fig. \ref{entangled_diffraction}(f). The simulation uses $\sinc^2$ functions on $x_+$ and $x_-$ with effective diffraction coefficients $a_+$ and $a_-$ as an envelope for the coincidence probability maps of Eq. (\ref{pDBC}) and Eq. (\ref{pSBC}). For the DBC case, $a_+=1.13a$ and $a_-=0.52a$, where $a$ is the expected diffraction coefficient for a model that considers the amplitude constant over the length of a slit. For same branch coincidences, $a_+=a_-=0.29a$ was found by tracing over one photon of the pair and then proceeding with the numerical integration. We found the results to be in good agreement with the experimental maps.

A similar asymmetry is reported in \cite{Ostermeyer:2009p2580} for a photon pairs passing through a blazed grating. There, also, the asymmetry is related to the degree of entanglement between the photons.

\subsection{Near-separable state}

\begin{figure}[htbp]
\centering\includegraphics[width=\textwidth]{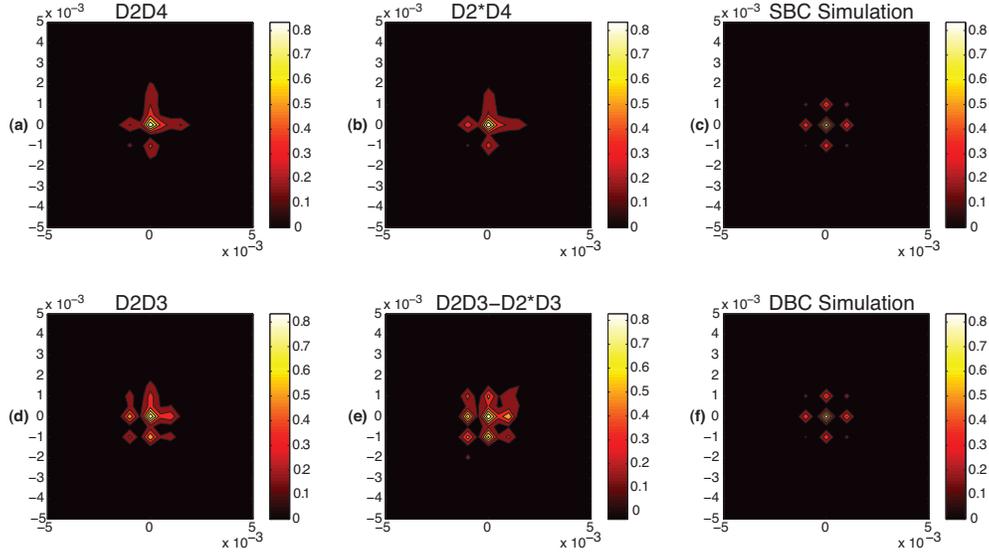}
\caption{Normalized two-photon measured and simulated broad scan coincidence maps for a four-photon path state of negligible spatial correlation. The upper row, from \textbf{(a)} to \textbf{(c)}, includes the measured coincidences on the H branch (D2D4), the calculated uncorrelated pattern corresponding to the product of the single counts (D2*D4), weighted by the repetition rate of the pump, and the SBC simulation, using Eq. (\ref{pSBC}) and Eq. (\ref{sincpm}). The lower row, from \textbf{(d)} to \textbf{(f)}, corresponds respectively to the coincidences between the H and V branches (D2D3) and to the same map corrected by subtracting the uncorrelated contribution (D2D3-D2*D3), as well as a simulated DBC map, using Eq. (\ref{pDBC}) and Eq. (\ref{sincpm}). The simulation parameters used are, as before, $\alpha_b^{exp}=\unit{86}{\degree}$ and $\varphi_b^{exp}=\unit{5}{\degree}$. For a measurement time of \unit{25}{s} per point, the maximum in map \textbf{(a)} corresponds to 75 coincidence counts and in map \textbf{(d)} to 117 coincidence counts.}
\label{diffraction_separable}
\end{figure}

In the near-separable case, $A(\xi_1,\xi_2)$ was indeed constant over the aperture function $T(\xi_1,\xi_2)$. Thus, we found by numerical integration of the equations represented in Eq. (\ref{sincpm}) that $a_+=a_-=0.99a$. This was expected, since the photons have almost no entanglement and, therefore, should behave as four independent photons diffracting through the slit aperture. It can be seen in Fig. \ref{diffraction_separable} that the envelopes for both SBC and DBC are very similar, with minor differences attributed in part to a small degree of entanglement present and also to image rendering artifacts.

\section{Discussion and conclusions}\label{discussion}
In this work we demonstrated experimental control of quantum correlations in a four-photon path state comprised of two pairs of photons. Our scheme adopted a high-efficiency type-II parametric down-conversion source coupled to a double slit by a variable linear optical setup, in order to obtain spatially encoded qubits. By altering experimental parameters, we have generated pairs with very high and very low degree of entanglement, and shown that the quantum correlations influence the two-photon interference pattern. That makes it possible to use a map of coincidence measurements for characterizing the four-photon path state. Moreover, we have shown that, in the case of type-II multi-pair generation, maps for coincidences between photons with the same polarization, as well as between photons with orthogonal polarization, can be used to obtain a full understanding of the phenomenon without having to resort to fourfold coincidence detection.

Finally, we observed that the diffraction pattern is also affected by the degree of transverse quantum correlation between the two photons of a pair at the double-slit plane. We were able to present a semi-quantitative approach that resulted in simulated maps in good agreement to the experimental results. A more detailed study aiming for a closed theoretical description of how entanglement plays a part in distorting the diffraction patterns observed on coincidence measurements is under way.

Currently, the main impediment for more efficient multi-pair experiments in quantum imaging is the fact that the area integrated by the detector is very small. The fast improvement on the field of CCD detectors, both in readout time and in sensitivity, opens the possibility for an experiment with four or more multi-pixel detectors with largely increased genuine multi-pair detection signal \cite{Abouraddy:2001p1710}.

Another direction of improvement is the expansion of the dimension of the logical basis from qubits to qudits. This can be implemented by increasing the number of slits and adjusting the quantum correlation control setup in an appropriate way. With a larger space, many interesting options become available, such as non-locality \cite{Collins:2002p040404,Thew:2004p010503,Vertesi:2010p060401} and contextuality \cite{Cabello:2010} tests.

\section*{Acknowledgements}
The authors would like to thank CNPq, CAPES and FAPEMIG Brazilian funding agencies, as well as the INCT-IQ (Instituto Nacional de Ciência e Tecnologia para Informação Quântica) for the financial support. P.-L. would like to thank Prof. Nivaldo Speziali for providing the double slits used in this work. We acknowledge support from Italian-Brazilian (CNR-CNPq) Contract (Quantum information in a high dimension Hilbert space).
\end{document}